\documentclass[aps,prd,nofootinbib,twocolumn,superscriptaddress,preprintnumbers,balancelastpage,longbibliography]{revtex4-1}

\usepackage{aas_macros}
\usepackage{titletoc}

\usepackage{placeins}
\usepackage{amsmath,amssymb,mathtools,bm}
\usepackage{graphicx, color, hepunits}
\usepackage[dvipsnames]{xcolor}
\usepackage{cancel}
\usepackage[normalem]{ulem}
\usepackage{float}
\usepackage{filecontents}
\usepackage{multirow}
\usepackage{tikz}
\usepackage{slashed}
\usetikzlibrary{decorations.pathmorphing}
\usetikzlibrary{arrows}
\usetikzlibrary{shapes.misc}
\usetikzlibrary{positioning}

 \usepackage{hyperref} 
\hypersetup{
    colorlinks=true,       
    linkcolor=blue,        
    citecolor=blue,        
    filecolor=magenta,     
    urlcolor=blue          
}
\usepackage[utf8]{inputenc}
\usepackage[english]{babel}
\let\vec\mathbf

\newcommand{\cmt}[1]{}

\begin{document}



\title{Multimessenger search strategy for composite dark matter with white dwarf data and gravitational wave detectors}


\author{Siyu Jiang}
\author{Aidi Yang}
\author{Fa Peng Huang}
\email{Corresponding Author.  huangfp8@sysu.edu.cn}
\affiliation{MOE Key Laboratory of TianQin Mission, TianQin Research Center for Gravitational Physics \& School of Physics and Astronomy, Frontiers Science Center for TianQin, Gravitational Wave Research Center of CNSA, Sun Yat-sen University (Zhuhai Campus), Zhuhai 519082, China}

\begin{abstract}
The nature of dark matter (DM) remains one of the most challenges in modern physics. Composite or macroscopic DM present a compelling alternative to conventional particle DM, yet their terrestrial search is notoriously challenging due to low number density.
This Letter presents a unified, multi-messenger search strategy for composite DM, dramatically improving existing constraints and proposing a new detection method.
We first perform a model-independent update of astrophysical constraints from compact objects, utilizing systematic calculations and additional white dwarf data related to ignition and subsequent supernovae. Crucially, for the first time, we explore novel signals of composite DM in future gravitational wave detectors like LISA, TianQin, and Taiji, performing detailed signal-to-noise ratio and Fisher matrix analyses.
We demonstrate that these detectors will possess the requisite sensitivity to probe untouched regions of the DM parameter space.
Our results underscore the unique power of the multi-messenger paradigm—spanning stellar astrophysics and gravitational wave astronomy—to explore this distinct and challenging frontier of DM physics. Our analyses extend broadly to a wide range of macroscopic or composite DM scenarios.
\end{abstract}

\maketitle

The particle nature of dark matter (DM) remains elusive, presenting one of the most enduring puzzles in fundamental physics. While experimental efforts have long focused on elementary particles, the absence of conclusive signals has reinvigorated interest in macroscopic composite states. Analogous to the formation of nuclei in the visible sector, dark sectors may aggregate into composite objects---such as Q-balls or Fermi-balls---arising from mechanisms like the Affleck-Dine scenario~\cite{Kusenko:1997zq,Kusenko:1997si} or first-order phase transitions in the early universe~\cite{Witten:1984rs,Krylov:2013qe,Huang:2017kzu,Hong:2020est,Jiang:2024zrb,Xie:2024mxr}. However, probing this macroscopic or composite frontier presents a distinct experimental challenge: as the constituent mass increases, the number density drops precipitously, rendering terrestrial direct detection experiments effectively blind to the resulting low flux.

To overcome this ``flux barrier'', one can utilize astrophysical scales. Compact objects like white dwarfs (WDs), acting as vast, long-exposure detectors, offer a powerful alternative~\cite{Jacobs:2014yca,Dvorkin:2013cea,SinghSidhu:2019znk,SinghSidhu:2019cpq,Badurina:2025xwl,Bai:2023lyf,Wadekar:2022ymq}. A composite or macroscopic DM object traversing a WD can dissipate sufficient kinetic energy via scattering to ignite a local thermonuclear runaway, potentially triggering a Type Ia supernova~\cite{Graham:2015apa,Graham:2018efk,SinghSidhu:2019tbr,Raj:2023azx}. Concurrently, lighter macroscopic DM candidates ($m_{\mathrm{DM}} \sim 1\text{--}10^6~\mathrm{GeV}$) are constrained by capture in neutron stars (NSs)~\cite{Bertone:2007ae,McCullough:2010ai,Hooper:2010es,Bertoni:2013bsa,Bramante:2015cua,Dasgupta:2019juq,Bell:2020jou,Dasgupta:2020dik,Bose:2022ola,Nguyen:2022zwb,Su:2024flx,Bhattacharjee:2025iip}, which can lead to heating or gravitational collapse~\cite{deLavallaz:2010wp,Horowitz:2020axx,Ema:2024wqr,McDermott:2011jp,Kouvaris:2011fi,Bramante:2013hn,Bell:2013xk,Janish:2019nkk,Lu:2024kiz,Liu:2024qbe,Dasgupta:2020mqg,Bhattacharya:2023stq,Kouvaris:2014rja}. DM inside NSs changes their equation of state, affecting mass, radius, and tidal deformability, which could be imprinted in gravitational wave (GW) signals from binary mergers~\cite{Mukhopadhyay:2015xhs,Mariani:2023wtv}.

Furthermore, DM yields unique signatures in GW experiments~\cite{adams2004directdetectiondarkmatter,Vinet:2006fj,Hall:2016usm,Grote:2019uvn,Jaeckel:2020mqa,Thoss:2025yht,Pierce:2018xmy,Morisaki:2018htj,Yu:2023iog,Yao:2024hap,Gue:2024txz,yao2025axionlikedarkmattersearch,xu2025distinguishingmonochromaticsignalslisa,gué2025discriminatingscalarultralightdark,Bhattacharya:2024pmp}, like LISA~\cite{amaroseoane2017laserinterferometerspaceantenna}, TianQin~\cite{TianQin:2015yph} and Taiji~\cite{Hu:2017mde}. For a review, see~\cite{miller2025gravitationalwaveprobesparticle}. Composite DM could interact directly with the satellites~\cite{adams2004directdetectiondarkmatter,Vinet:2006fj,Hall:2016usm,Grote:2019uvn,Jaeckel:2020mqa,Thoss:2025yht}, or affect the photon propagation through the Shapiro effect~\cite{Siegel:2007fz,Baghram:2011is,Lee:2020wfn,Lee:2021zqw,Ramani:2020hdo,Du:2023dhk}.

In this Letter, we place composite DM ``under siege'' by establishing a unified framework that bridges the gap between astrophysical observation and precision interferometry. Focusing on composite DM with an extra long-range Yukawa attraction, we first derive updated constraints from WD ignition, incorporating refined stellar density profiles and expanded data of existing WDs. Using the updated bounds, we then demonstrate the capability of LISA, TianQin, and Taiji to detect DM with non-gravitational interactions, utilizing detailed signal-to-noise ratio and Fisher matrix analyses. The conclusions can be generically applied to a broad class of macroscopic or composite DM scenarios.

{\bf Composite DM with long-range interaction}---For various composite DM models, the DM is composite of dark fermions $X$. We show three representative models below for a broad class of composite DM models, where the DM mass can be parameterized as 
\begin{equation}\label{DMscaling}
    M_{\mathrm{DM}} = \varepsilon \times \frac{4\pi}{3}R_{\mathrm{DM}}^3
\end{equation}
where $\varepsilon$ is the density of the constituents of DM. 

(a) Dark quark nuggets (DQNs): DQNs are made of dark quarks that are in the deconfined phase of the underlying dark
quantum chromodynamics (QCD)~\cite{Bai:2018dxf}. They are formed during the dark QCD first-order phase transition and are stabilized by the balance of the Fermi pressure of the dark quarks $\frac{N_dN_f\mu_{\mathrm{eq}}^4}{12\pi^2}$ and the vacuum pressure $\Delta P_{\text{vacuum}} = \Lambda_d^4$, where we have assumed dark quarks have $N_f$ flavors and $N_d$ colors.  $\mu_{\mathrm{eq}} = (24\pi^2\Delta P_{\text{vacuum}}/g_{\mathrm{dof}})^{1/4}$ is the equilibrium chemical potential with $g_{\mathrm{dof}} = 2N_dN_f$ being the degree of freedom of dark quarks and $\Lambda_d$ is the confinement scale. For DQNs,
\begin{equation}
    \varepsilon = 4 \Lambda_d^4
\end{equation}
The number density of constituents $X$ is $n_X= \frac{g_{\mathrm{dof}}}{6\pi^2}\mu_{\mathrm{eq}}^3 $.

(b) Axion quark nuggets (AQNs): AQNs could form from $N=1$ QCD axion domain wall bubbles~\cite{Zhitnitsky:2002qa}. During the quark-hadron transition (at $T\sim 170~\mathrm{MeV}$ ), some of these closed bubbles avoid collapse by accreting quarks, thereby giving rise to AQNs.
The interior of an AQN is in a color-superconducting phase, while the exterior is in the hadronic phase. The external pressure from the axion domain wall balances the internal Fermi pressure, stabilizing the AQN. Different from Witten's idea, AQNs form without the requirement of first-order phase transition, and are stabilized by the axion domain wall. We define the AQN mass by using the nuclear mass density~\cite{Majidi:2024mty},
\begin{equation}
    \varepsilon = \rho_n = 3.5\times 10^{14}~\mathrm{g}~\mathrm{cm}^{-3}
\end{equation}

(c) Fermi-balls: Similar to the DQNs, Fermi-balls can be formed during a cosmic first-order phase transition, and are stabilized by the balance of the vacuum energy difference $\Delta V$ and the Fermi degenerate pressure of the dark fermions~\cite{Huang:2017kzu,Hong:2020est}.  The energy of a Fermi-ball with global charge $N_{\mathrm{FB}}$ is $E \simeq \frac{g_{\mathrm{dof}}}{6\pi} \left( \frac{9\pi}{2g_{\mathrm{dof}}} \right)^{4/3} \frac{N_{\text{FB}}^{4/3}}{R} + \frac{4\pi}{3} \Delta V R^3$ with $N_{\mathrm{FB}}=\frac{2g_{\mathrm{dof}}}{9\pi}\mu_{}^3 R_{}^3$~\cite{Gresham:2017zqi}. After minimizing $E$ with respect to the radius at $R= R_{\mathrm{FB}}$, one can get $\mu = \mu_{\mathrm{eq}} = (24\pi^2\Delta V/g_{\mathrm{dof}})^{1/4}$ and the Fermi-ball mass can be expressed by the formula of Eq.~\eqref{DMscaling} with
\begin{equation}
    \varepsilon = 4 \Delta V \simeq 4 w^4
\end{equation}
where $w$ is the vacuum expectation value of the phase transition potential.

In summary, composite DM formed by the condensation of fermions have some common characteristics: they both have the mass-radius relation \eqref{DMscaling} and 
\begin{equation}
    \mu_{\mathrm{eq}} = \left(\frac{6\pi^2 \varepsilon}{g_{\mathrm{dof}}}\right)^{1/4},\quad N_{\mathrm{DM}}=\frac{2g_{\mathrm{dof}}}{9\pi}\mu_{\mathrm{eq}}^3 R_{\mathrm{DM}}^3
\end{equation}

Composite DM could interact with ordinary matter via a light scalar, one simple case can be seen in Ref.~\cite{Acevedo:2021kly}.
We introduce a
long-range scalar mediator that couples the DM via $y_{\mathrm{DM}}\bar{X} \phi X$ and the nucleons via $y_{\mathrm{SM}} \bar{n} \phi n$, such that an additional Yukawa attractive potential arises between two objects $i,j \in \{\mathrm{SM}, \mathrm{DM}\}$, 
\begin{equation}\label{Yukawa}
V_{i-j}=-M_i M_j \frac{G}{r}\left(1+\delta_i \delta_j e^{-m_{\mathrm{med}} r}\right)\,\,,
\end{equation}
with $\delta_i = \frac{y_i N_i }{\sqrt{4\pi G}M_i}$. For nucleons $\delta_{\mathrm{SM}} = \frac{y_{\mathrm{SM}}}{\sqrt{4\pi G}\bar{m}_n}$ with $\bar{m}_n$ being nucleon mass. The fermion condensate will induce an effective potential of the scalar inside DM~\cite{Bai:2023lyf},
\begin{equation}
    V_{\mathrm{eff}}(\phi) \approx \frac{g_{\mathrm{dof}}}{8\pi^2} \mu^2\left[\mu^2-(m_X-y_{\mathrm{DM}}\phi)^2 \right]
\end{equation}
where $m_X$ is the fermion mass outside the DM and generally has the value of $m_X \simeq \varepsilon^{1/4}$ in the phase transition model. The scalar mass $m_{\mathrm{in}}= \frac{g_{\mathrm{dof}}^{1/2}y_{\mathrm{DM}}\mu_{\mathrm{eq}}}{2\pi}$ inside the composite DM could be much larger than the scalar mass $m_{\mathrm{med}}$ outside DM. The $m_{\mathrm{in}}$ could also be dramatically modified by including the scalar self-interactions~\cite{Lu:2024xnb}. The term linear in $\phi$, proportional to $y_{\mathrm{DM}}m_X$, gives rise to the effective charge density $N_{\mathrm{DM}}^{\mathrm{eff}} = \frac{g_{\mathrm{dof}}}{4\pi^2} m_X \mu_{\mathrm{eq}}^2$ that is capable of emitting the scalar field. If the effective mediator mass within the DQN exceeds 
$1/R_{\mathrm{DM}}$, the dark quark plasma will induce screening effects, thereby suppressing the DQN's effective $\phi$-charge by an additional factor of $1/(m_{\mathrm{in}}R_{\mathrm{DM}})^2$~\cite{Bai:2023lyf}. 
We will assume the force range $\lambda = m_{\mathrm{med}}^{-1}$ is larger than the typical lengths in this work, i.e., the radius of the WD or NS, and the arm length of the GW detector; for instance, we take $\lambda=0.1~\mathrm{pc}$. Then the attractive potential between the DM and the nucleons is simply obtained by replacing the Newton constant in the gravitational potential by $G \rightarrow G(1+\tilde{\alpha})$ with $\tilde{\alpha} = \delta_{\mathrm{SM}} \delta_\mathrm{DM}$.
The strength of the long-rang Yukawa interaction between SM particles is constrained by the MICROSCOPE mission’s weak equivalence principle test. The MICROSCOPE experiment measures the E\"otv\"os parameter by comparing the free-fall accelerations of two test masses of different compositions (titanium and platinum alloys) on board a satellite. If a fifth force, coupling to quantities such as baryon number $B$ or lepton number $L$, it would cause a composition-dependent differential acceleration, allowing the experiment to set upper limits on the strength of such a new force relative to gravity. While $\left|\delta_{\mathrm{SM}}\right|<3 \times 10^{-6}$ has been strongly constrained by the MICROSCOPE experiment~\cite{Berge:2017ovy,Fayet:2018cjy}, the DM coupling could be several orders larger, since the only relevant experimental constraints are imposed by the influence of DM self-interaction on structure formation~\cite{Spergel:1999mh,Kahlhoefer:2013dca,Harvey:2015hha}. Observation of the bullet cluster constrained the cross-section of DM to be $\sigma_{\mathrm{DM}-\mathrm{DM}}/M_\mathrm{DM} \lesssim 1 \mathrm{~cm}^2 / \mathrm{g}$, from which the constraint on $\delta_{\mathrm{DM}}$ is given by~\cite{Gresham:2022biw}:
\begin{equation}\label{bullet}
\delta_{\mathrm{DM}}^2 <2\times10^{6}\left(\frac{M_{\mathrm{DM}}}{M_{\odot}}\right)^{-1/2}\left(\frac{v_{\mathrm{DM}}}{10^{-2}}\right)^{2}e^{4\times10^{-3}\sqrt{\frac{M_{\mathrm{DM}}}{M_{\odot}}}\frac{\mathrm{pc}}{\lambda}}\,\,,
\end{equation}
with $v_{\mathrm{DM}}$ being the DM velocity. The composite DM could collapse into a black hole when its radius drops below the Schwarzschild radius $2G M_{\mathrm{DM}}$.

{\bf Runaway fusion of compact objects}---
Firstly, we briefly discuss the condition of triggering runaway fusion in a WD core and NS ocean. When DM passes a WD or NS, it may deposit energy through scattering with  C or O ions. If the energy deposition $E_{\mathrm{dep }}$ is enough for the ions to overcome their mutual Coulomb barrier, the runaway fusion occurs. And for a C/O white dwarf, there are three primary fusion processes: ${}^{12}\mathrm{C}+{}^{12}\mathrm{C}$, ${}^{12}\mathrm{C}+{}^{16}\mathrm{O}$ and ${}^{16}\mathrm{O}+{}^{16}\mathrm{O}$.  Therefore, the first condition for fusion is that the energy deposition must raise the temperature of a certain area above some critical value $T_{\mathrm{crit}} \sim \mathrm{MeV}$. Here $T_{\mathrm{crit}}$ is set to be $0.5~\mathrm{MeV}$~\cite{1992ApJ...396..649T}. The second condition is that the timescale for cooling due to thermal diffusion has to be longer than the timescale corresponding to the heating due to fusion. The former increases with the size of the heated region but the latter is independent of the size of the region; therefore, there is a critical size $\lambda_{T}$ beyond which the heat cannot be efficiently diffused~\cite{Graham:2018efk}. Following discussions in \cite{Fedderke:2019jur}, the ignition condition reads
\begin{equation}\label{firstcondition}
E_{\mathrm{dep }} \geq E_{\mathrm{trig}} \equiv \frac{4\pi}{3}\rho_{}\lambda_{T}^3\left(\rho_{}, T_{\text {crit }}\right) \bar{c}_p\left(\rho, T_{\text {crit }}\right) T_{\text {crit }}\,\,,
\end{equation}
where $\rho_{}$ is the local density of WDs or NSs. $\bar{c}_p \simeq c_p^{\mathrm{ion}} / 2+c_p^\gamma / 4+c_p^{e}$ is the averaged heat capacity with the heat capacity for ions, photons, and electrons (see Supplementary Material).

It is often assumed that DM triggers the ignition near the center of a CO WD. Obtaining the precise density profile of WDs requires solving the Tolman–Oppenheimer–Volkoff (TOV) equation~\cite{Tolman:1939jz,Oppenheimer:1939ne}. We use the analytic fit of the WD-mass
 central density relationship introduced by Ref.~\cite{Fedderke:2019jur},
\begin{equation}
1+\left(\frac{\rho_{\mathrm{WD}}}{1.95 \times 10^6 \mathrm{~g} / \mathrm{cm}^3}\right)^{\frac{2}{3}} \approx\left[\sum_{i=0}^6 c_i\left(\frac{M_{\mathrm{WD}}}{M_{\odot}}\right)^i\right]^{-2}
\end{equation}
with $\left\{c_i\right\}$ =\{$1.0033$, $-0.3087$, $-1.1652$, $2.0211$, $-2.0604$, $1.1687$, $-0.2810$\}.

The trigger length $\lambda_T$ is obtained by comparing the nuclear energy generation rate and the heat diffusion rate. However, the specific nuclear energy generation rates of  the WD generally require numerical simulation, and the Ref.~\cite{Fedderke:2019jur} provides an analytical scaling relation,
\begin{align}\label{scaling}
\lambda_{T}= \begin{cases}\lambda_1^{}\left(\frac{\rho}{\rho_1}\right)^{-2},~\rho \leq \rho_1\,\,, \\ \lambda_1^{}\left(\frac{\rho}{\rho_1}\right)^{\ln \left(\lambda_2^{} / \lambda_1^{}\right) / \ln \left(\rho_2 / \rho_1\right)},~\rho_1<\rho \leq \rho_2\end{cases}
\end{align}
with $\left\{\lambda_1^{}, \lambda_2^{}\right\}=\left\{1.3 \times 10^{-4} \mathrm{~cm}, 2.5 \times\right. \left.10^{-5} \mathrm{~cm}\right\}$ and $\left\{\rho_1, \rho_2\right\}=\left\{2 \times 10^8 \mathrm{~g}\cdot \mathrm{cm}^{-3}, 10^{10} \mathrm{~g} \cdot \mathrm{cm}^{-3}\right\}$ for WDs. We take 4361 WDs in Montreal White Dwarf
Database~\cite{dufour2016montrealwhitedwarfdatabase} with known lifetimes and masses spanning $0.8<M_{\mathrm{WD}}/M_\odot<1.4$. The minimum trigger length is found to be $\lambda_T \approx 4.8  \times 10^{-5}~\mathrm{cm}$.

Unlike WDs, the thermal nuclear runaway causes the NSs to undergo a superburst instead of a supernova. The outer layer of NS consists of an ocean of heavy elements, where the mass fraction of carbon is approximately 10\%~\cite{Cumming:2001wg}. For NSs,  the precise nature of the initiation of a superburst is still unclear. However, similar to the WDs, the trigger length is also derived by comparing the carbon fusion rate with the heat diffusion rate, and it is expected that the scaling relation \eqref{scaling} still holds for NSs~\cite{SinghSidhu:2019tbr}. For the density profile of NSs, we adopt the Tolman VII solution~\cite{PhysRev.55.364}:
\begin{equation}\label{NSprofile}
    \rho(r) \simeq \rho_{\mathrm{NS}} (1-r^2/R_{\mathrm{NS}}^2)
\end{equation}
with $\rho_{\mathrm{NS}} = 15M_{\mathrm{NS}}/(8\pi R_{\mathrm{NS}}^3)$ which is derived by using $4\pi \int_{0}^{R_{\mathrm{NS}}} \rho(r) r^2 dr = M_{\mathrm{NS}}$. 
Improved analytic profile of NS is shown in Ref.~\cite{Jiang:2019vmf} and numerical density profile for different equations of state is given in Ref.~\cite{crustdensityprofileRRIIIA}. We integrate the density profile \eqref{NSprofile} to derive the position corresponding to the maximum ignition column depth of $4\times10^{13}~ \mathrm{g}\cdot\mathrm{cm}^{-2}$ presented in Fig. 1 of Ref.~\cite{Cumming:2001wg}. The superbursts occur via pure C burning and for NS oceans the carbon mass fraction is about 10\%.

The superbursts source 4U 1820-30 is used to place constraints on the composite DM. The mass and radius of the NS are~\cite{Guver:2010td}
$
M_{\mathrm{NS}} = 1.58~M_{\odot}$ and $R_{\mathrm{NS}} = 9.11~\mathrm{km}
$ respectively.
The recurrence time of 4U 1820-30 is taken to be $t_{\mathrm{rec}}=2.5~\mathrm{yr}$~\cite{SinghSidhu:2019tbr} and its Galactic position is 1.2 kpc. The trigger length of NS is derived by integrating the neutron star crust density profile in Ref.~\cite{crustdensityprofileRRIIIA} to obtain the crustal column depth and then comparing it to the maximum ignition column depth of $4\times10^{13}~ \mathrm{g}\cdot\mathrm{cm}^{-2}$ presented in Fig. 1 of Ref.~\cite{Cumming:2001wg}.

As the composite DM passes through a compact object with mass $M_{\mathrm{CO}}$ and radius $R_{\mathrm{CO}}$, it can deposit its energy through scattering. The discussions for elastic and inelastic scattering between DM and nuclear can be found in Ref.~\cite{HoefkenZink:2024hor}, and it has been shown that the elastic scattering dominates when the kinetic energy of DM constituents is lower than the nucleon mass. A related discussion of energy deposition in molecular clouds can be found in Ref.~\cite{AngelesPerez-Garcia:2013fpo}.
The  energy transfer reads,
\begin{equation}
    \frac{dE_{\mathrm{dep}}}{dx} \approx \rho \sigma_{\mathrm{DM}-A} \tilde{v}_{\mathrm{DM}}^2\,\,,
\end{equation}
where $\tilde{v}_{\mathrm{DM}} \simeq v_{\mathrm{esc}} \sqrt{1+\tilde{\alpha}}$ is the DM velocity with $v_{\mathrm{esc}} = \sqrt{2GM_{\mathrm{CO}}/R_{\mathrm{CO}}}$ being the escape speed, and $\sigma_{\mathrm{DM}-A}$ is the scattering cross section between DM and carbon ions. As shown in the Ref.~\cite{Ponton:2019hux}, when the DM's radius is large enough, and its interior is in the symmetry unbroken phase, the nucleon entering the DM immediately feels a strong potential well and is almost certain to be scattered or absorbed. Therefore, for a large-radius composite DM, it can be viewed as a ``hard sphere" and its scattering cross section is primarily determined by its geometric size, i.e., $\sigma_{\mathrm{geo}} \approx \pi R_{\mathrm{DM}}^2$. Besides, 
due to the extra long range interaction, the cross section caused by the $\phi$-exchange reads~\cite{Grabowska:2018lnd,Xu:2020qjk,Bhoonah:2020dzs,Acevedo:2021kly},
\begin{equation}\label{crosssection}
  \sigma_{\phi} \simeq  \frac{16\pi G^2 \tilde{\alpha}^2 M_{\mathrm{DM}}^2 \bar{m}_A^4 |F_A(q)|^2 |F_{\mathrm{DM}}(q)|^2}{(q^2+m_\phi^2)^2\times \max(1,m_{\mathrm{in}}^2R_{\mathrm{DM}}^2)} 
\end{equation}
where $\bar{m}_A = A\bar{m}_n$ is the
nuclear mass. 
Here,
we have used the fact $M_{\mathrm{DM}} \gg \bar{m}_A$ so that the reduced mass $\mu_{\mathrm{re}} = \frac{M_{\mathrm{DM}} \bar{m}_A}{M_{\mathrm{DM}} +\bar{m}_A} \approx \bar{m}_A$. $q \simeq \sqrt{2\bar{m}_A E_{\mathrm{nr}}} $ is the average momentum transfer between DM and nucleus, and $E_{\mathrm{nr}} = \bar{m}_A \tilde{v}_{\mathrm{DM}}^2$. The Helm nucleus form factor $F_A(q)$ is given by~\cite{Helm:1956zz,Lewin:1995rx} 
\begin{equation}
    F_A(q) = \frac{3j_1(qR_A)}{qR_A} e^{-q^2 R_A^2}\,\,,
\end{equation}
with $R_A \simeq (1.25~\mathrm{fm})A^{1/3}$ being the nuclear radius and $j_1$ the
Bessel function of the first kind. In order to take into account the structure of composite DM, we take the DM form factor to be
\begin{equation}
    F_{\mathrm{DM}}(q) = \frac{3j_1(qR_{\mathrm{DM}})}{qR_{\mathrm{DM}}}\,\,.
\end{equation}
Summarily, the cross section between DM and nuclear is 
\begin{equation}
    \sigma_{\mathrm{DM}-A} = \max(\sigma_{\mathrm{geo}}, \sigma_{\phi})
\end{equation}Given that $R_{\mathrm{DM}} \gg \lambda_T$, we have to multiply $E_{\mathrm{dep}} $ by $\left(\lambda_T/R_{\mathrm{DM}} \right)^3$ after taking into account the energy deposition in the critical volume, $\frac{4\pi}{3}\lambda_T^3$. Then the energy deposition to trigger the ignition reads,
\begin{equation}
    \frac{dE_{\mathrm{dep}} }{dx}\gtrsim \frac{E_{\mathrm{trig}}}{\lambda_T} \times \max\left[1, \left(\frac{R_{\mathrm{DM}}}{\lambda_T}\right)^2\right]\,\,.
\end{equation}

{\bf Astrophysical constraints from supernova and superburst.}---
In addition to the ignition, it is important to investigate the event rate at WDs/NSs. The encounter rate between DM and WD/NS relies on the maximum impact factor $b_{\mathrm{max}}$~\cite{Gresham:2022biw},
\begin{equation}
    \Gamma_{\mathrm{enc}}^{\mathrm{DM}}(r)=f_{\mathrm{DM}}\frac{\rho_{\mathrm{DM}}(r)}{M_{\mathrm{DM}}}v_{\mathrm{DM}}(r)\pi b_{\mathrm{max}}^2\,\,.
\end{equation}
Complete formula of $b_{\mathrm{max}}$ can be found in Supplementary Material. $f_{\mathrm{DM}}$ is the fraction of composite DM making up the whole DM. This is phenomenologically reasonable in the theory of DM formation during a first-order phase transition~\cite{Hong:2020est}. 
For WDs, DM is chosen to be at the solar position, $r_\odot=8.3~\mathrm{kpc}$, where $\rho_{\mathrm{DM}}(r_{\odot})=0.4\mathrm{~GeV/cm}^{3}$ and $v_{\odot} \approx 220 \mathrm{~km} / \mathrm{s}$~\cite{10.1093/mnras/221.4.1023,Drukier:1986tm,Bland-Hawthorn:2016lwg,Evans:2018bqy}. The expected number of DM passing through the WDs is
\begin{equation}
N_{\mathrm{enc}}^{\mathrm{WD}}=\sum_{i}\Gamma_{\mathrm{enc}}^{\mathrm{DM}}\tau_{\mathrm{WD},i}\,\,.
\end{equation}
We set $N_{\mathrm{enc}}^{\mathrm{WD}}=3$ which may be ruled out at 95\% confidence, as pointed in Ref.~\cite{SinghSidhu:2019tbr}. For NSs, the $\tau_{\mathrm{WD}}$ has to be replaced by $t_{\mathrm{rec}}$. To evaluate the DM velocity at the position of NS, we use the Maxwell-Boltzmann dispersion speed $v_{\mathrm{DM}}(r) = \sqrt{3GM_{}(r)/2r}$ where $M_{}(r)$ is the DM mass enclosed in $r$~\cite{Raj:2023azx}. By adopting the Navarro-Frenk-White profile~\cite{Navarro:1996gj,Akita:2022lit}, 
then $M(r) =\int_0^r d^3 \vec{r}^{\prime} \rho_{\mathrm{DM}}\left(r^{\prime}\right) $. As more DM is focused, constraints are amplified when introducing the Yukawa interaction. The maximum value of DM mass to be constrained by WDs is $6.25\times 10^{18}~\mathrm{g}$ and $6.46\times 10^{20}~\mathrm{g}$ for $\tilde{\alpha} = 0$ and $\tilde{\alpha}=100$, respectively. 

For WDs, there will be a surface above which the electrons become non-degenerate, and there the density $\rho_{\mathrm{env}}$ is typically only about 0.1\% of the central density. The temperature $T_*$ at the transition point can be determined by equating the electron pressure of the degenerate layer to that of the non-degenerate layer~\cite{Shapiro:1983du}. More details can be found in the Supplementary Material. Then the radius $R_*$ at which $T=T_*$ can be obtained from
\begin{equation}
    T_*=\frac{1}{4.25}\frac{\bar{\mu} \mu_a}{k}\frac{GM_{\mathrm{WD}}}{R_{\mathrm{WD}}}\left(\frac{R_{\mathrm{WD}}}{R_*}-1\right)\,\,,
\end{equation}
where $\mu_a$ is the atomic mass unit. And $\bar{\mu}\equiv \left(\sum_i X_i (1+Z_i) / A_i\right)^{-1}$ is the mean molecular weight with $X_i$, $Z_i$, and $A_i$ being the mass fraction, charge, and atomic mass number of ion species $i$, respectively. 
The width of a non-degenerate WD envelope is defined by $R_{\mathrm{env}} = R_{\mathrm{WD}} - R_*$.
The condition for DM to penetrate into the WDs is that the energy loss of DM cannot exceed the incident kinetic energy of DM,
\begin{equation}
    \frac{dE_{\mathrm{dep}}}{dx} \approx \rho_{\mathrm{env}} \sigma_{\mathrm{DM}-A} \tilde{v}_{\mathrm{DM}}^2 < \frac{M_{\mathrm{DM}} \tilde{v}_{\mathrm{DM}}^2}{R_{\mathrm{env}}}\,\,.
\end{equation}
From which one can obtain $\frac{\sigma_{\mathrm{DM}-A}}{M_{\mathrm{DM}}} < \frac{1}{\rho_{\mathrm{env}}R_{\mathrm{env}}}$. We take $\rho_{\mathrm{env}}$ to be $10^{-3}$ of the central density, $\rho_{\mathrm{env}} = 10^{-3} \rho_{\mathrm{WD}}$. From the updated WD database and the rigorous analyses above we obtain a conservative bound $\sigma_{\mathrm{DM}-A}/M_{\mathrm{DM}} > 7.3\times 10^{-11}~\mathrm{cm}^2/\mathrm{g}$. This value is about 100 times larger than that in \cite{Graham:2018efk} and $10^6$ times larger than the value reported in \cite{SinghSidhu:2019tbr}, but it is 300 times smaller than the value in \cite{Raj:2023azx}.

For NS crusts, the condition that DM crosses the non-degenerate layer is set by using the maximum ignition column depth at the carbon ocean in Ref.~\cite{Cumming:2001wg}, which is about $\rho_{\mathrm{env}}R_{\mathrm{env}} \simeq 4\times10^{13}~ \mathrm{g}\cdot\mathrm{cm}^{-2}$.

{\bf Signals at GW experiments.}---
As a composite DM object passes near the GW detection satellite, its attractive influence causes differential acceleration among the test masses of GW detectors, producing a detectable Doppler signal. We assume the process occurs within the $x-y$ plane, with the DM moving along the $y$-direction at velocity $v_{\mathrm{DM}}$. The resulting differential acceleration of the nearest interferometer node can be expressed as~\cite{Vinet:2006fj,Jaeckel:2020mqa}
\begin{equation}
    \vec{g}(t)=\frac{\tilde{G}M_{\mathrm{DM}}}{D^2}\frac{1}{\left[1+\left(\frac{v_{\mathrm{DM}}t}{D}\right)^2\right]^{3/2}}
\begin{pmatrix}
1 \\
\frac{v_{\mathrm{DM}}t}{D} \\
0
\end{pmatrix}\,\,,
\end{equation}
where $D$ is the impact factor of DM with respect to the nearest satellite and   $\tilde{G}$ is given by $G(1+\tilde{\alpha})$ after taking into account the long-range Yukawa interaction.
In the following, we use the SI units to make the results clearer. After integration, the velocity of the test mass induced by the trajectory of DM reads,
\begin{equation}\label{vt}
    \vec{v}(t, D)=\frac{\tilde{G}M_{\mathrm{DM}}}{Dv_{\mathrm{DM}}}
\begin{pmatrix}
1+\frac{v_{\mathrm{DM}}t/D}{\sqrt{1+v_{\mathrm{DM}}^2t^2/D^2}} \\
-\frac{1}{\sqrt{1+v_{\mathrm{DM}}^2t^2/D^2}} \\
0
\end{pmatrix}\,\,.
\end{equation}
So in the limit $t\rightarrow -\infty$, the test mass is at rest. At $t\rightarrow \infty$, the velocity of the satellite approaches the $x$-direction and is equal to $\frac{2\tilde{G}M_{\mathrm{DM}}}{Dv_{\mathrm{DM}}}$. We approximate the impact factor between the DM and the other two nodes to be $D'=D+L$ by taking into account the larger distance with respect to the other two nodes due to the arm length. Some works consider the close approach limit, where the DM is restricted to passing by an interferometer node with an impact parameter smaller than the characteristic size of the detector, $D < L$~\cite{Vinet:2006fj,Jaeckel:2020mqa}.
We use $U_i$ and $V_i$ to encode the Doppler shifts of the laser beams that transmissions among detector nodes. In order to effectively cancel the laser noise,  time-delay interferometry is employed~\cite{Armstrong_1999,Dhurandhar:2001tct,Tinto:2020fcc}. For instance, the signal in Michelson-like X channel combination reads,
\begin{align}
    &X(t) = U_1(t) +V_1(t) -U_1\left(t-\frac{2L}{c}\right) - V_1\left(t-\frac{2L}{c}\right) + \nonumber\\
    &U_2\left(t-\frac{3L}{c}\right) + V_3\left(t-\frac{3L}{c}\right) - U_2\left(t-\frac{L}{c}\right)- V_3\left(t-\frac{L}{c}\right), \nonumber\\
& \quad \quad  \quad \quad U_1(t)=\vec{n}_2 \cdot \frac{\vec{v}_1(t)-\vec{v}_3(t-L / c)}{c}\,\,,\nonumber\\
&\quad \quad  \quad \quad V_1(t)=\vec{n}_3 \cdot \frac{\vec{v}_1(t)-\vec{v}_2(t-L / c)}{c}\,\,,
\end{align}
and permutation symmetry $1 \rightarrow 2 \rightarrow 3 \rightarrow 1$. $\vec{v}_i$ denotes the corresponding velocity perturbation of the $i$-th node induced by attractive interaction, while $\vec{n}_i$ refers to the unit vector between two nodes and the opposite node $i$. Note that $\vec{v}_1(t) = \vec{v}(t,D)$ and $\vec{v}_2(t) = \vec{v}_3(t) = \vec{v}(t,D')$. By using the relation $\vec{n}_2 + \vec{n}_3 = -\vec{n}_1$,
\begin{align}\label{Xt}
&X(t) = -\vec{n}_1 \cdot \left[\frac{\vec{v}(t, D) - \vec{v}(t - 4L/c, D)}{c} \right. \nonumber \\
& \left. + 2\frac{\vec{v}(t-3L/c, D')}{c} - 2\frac{\vec{v}(t-L/c,D')}{c}\right]\,\,.
\end{align}
For arbitrary orientation of the detector plane relative to the DM, $\vec{n}_1 = (\sin\vartheta\cos\varphi, \sin\vartheta\sin\varphi, \cos\vartheta)$.  
The GW detector is moving through the halo at a constant velocity, and the DM velocity follows the Maxwell-Boltzmann distribution with the central value $v_{\odot}$~\cite{Kamionkowski:1997xg}. We set $v_{\mathrm{DM}} \approx v_{\odot}$ for simplicity. Finally, the power spectral density (PSD) of the signal reads,
\begin{align}
P(\omega)  
=\frac{1}{4 \pi} \int \mathrm{d} \cos \vartheta \int \mathrm{~d} \varphi\left|\tilde{X}(\omega)\right|^2\,\,. 
\end{align}
where $\tilde{X}(\omega)$ is the Fourier transformation of $X(t)$ and $\omega = 2\pi f$. 
The complete formula for signal PSD in the three TDI channels and the corresponding noise PSD can be seen in the Supplementary Material.  

\begin{figure}[!ht]
    \centering
    \includegraphics[width=1\linewidth]{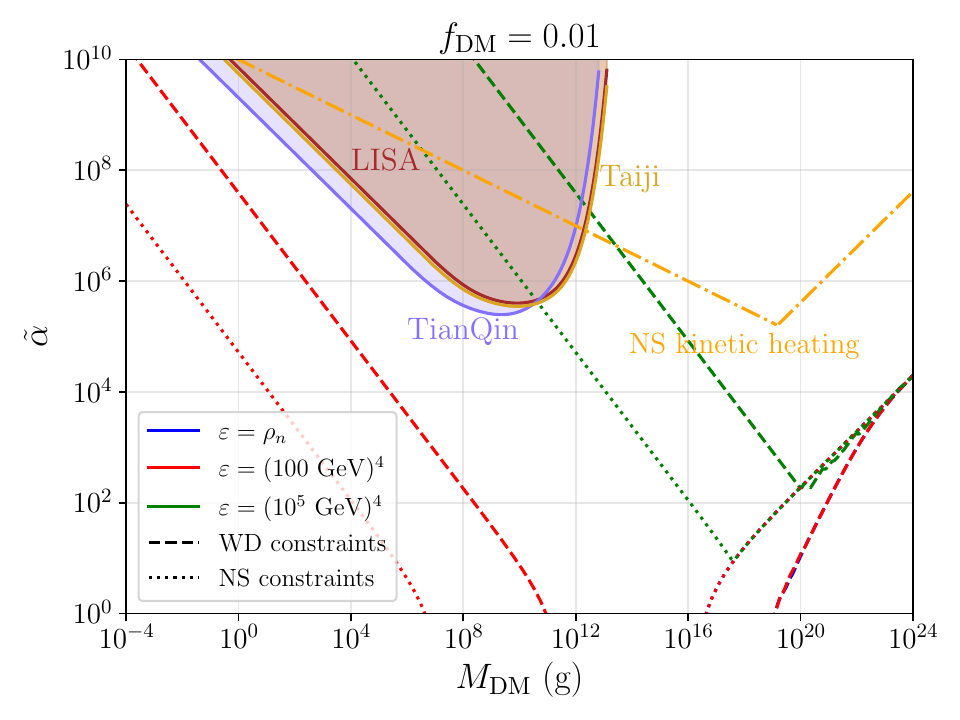}
    \caption{Sensitivity of GW detectors to composite DM with updated astrophysical constraints. The orange line denotes the constraints from NS kinetic heating.}
    \label{DM_alphaX_MX_001}
\end{figure}

We employ the X channel to derive our main result. The detectability of GW detectors is evaluated by the signal-to-noise ratio (SNR),
\begin{equation}
\mathrm{SNR} =\left(4\int_0^\infty\mathrm{d}\omega\frac{P(\omega)}{S_n(\omega)}\right)^{\frac{1}{2}} \,\,.
\end{equation}
where $S_n(\omega)$ is the noise PSD for LISA~\cite{amaroseoane2017laserinterferometerspaceantenna}, TianQin~\cite{TianQin:2015yph} and Taiji~\cite{Hu:2017mde}. To avoid the constraints of MICROSCOPE and the bullet cluster, $f_{\mathrm{DM}}$ and $\delta_{\mathrm{SM}}$ are set to be $f_{\mathrm{DM}}=0.01$ and $\delta_{\mathrm{SM}} = 10^{-6}$, respectively. The DM impacts on GW detectors are approximately a Poisson process. The corresponding probability is $( N_{\mathrm{enc}}^{\mathrm{GW}})^n\mathrm{e}^{- N_{\mathrm{enc}}^{\mathrm{GW}}}/n!$ where $n$ and $N_{\mathrm{enc}}^{\mathrm{GW}}=  \left(f_{\mathrm{DM}} \pi D^2 \rho_{\mathrm{DM}} v_{\mathrm{DM}}/M_{\mathrm{DM}}\right)T_{\mathrm{obs}}^i$ are the number of DM and the expected number of DM passing through the detector $i$, respectively. We restrict $N_{\mathrm{enc}}^{\mathrm{GW}}=1$ so that at least one DM induced event will be expected during the observation time $T_{\mathrm{obs}}^i$. The results of the sensitivity of the GW detectors after fixing $\mathrm{SNR}\geq 10$ are shown in Fig.~\ref{DM_alphaX_MX_001}. The detectability of GW detectors is strongest for DM with mass around $10^{11}-10^{12}~\mathrm{g}$. Below this value, the PSD of the signal is proportional to $\tilde{G}^2M_{\mathrm{DM}}^2$, so $\tilde{\alpha}$ must be increased to compensate for the decrease in $M_{\mathrm{DM}}$. On the other hand, if the DM mass is too large, the impact factor $D$ has to be increased for a fixed event rate. The signals suffer an extra exponential suppression, and then $\tilde{\alpha}$ needs to be increased again. The NS heating constraint due to the energy transfer from DM that is not gravitationally captured has been considered~\cite{Gresham:2022biw}. The composite DM suffers from the constraints of gravitational lensing~\cite{Croon:2020ouk,Niikura:2017zjd}, which disappears for $f_\mathrm{DM} \lesssim 0.01$.

In Fig.~\ref{DM_alphaX_MX_001}, we also show the constraints from WDs and NS, following the detailed procedures in the previous section. It can be seen that for composite DM with relatively low interior density, $\varepsilon \lesssim (100~{\mathrm{GeV}})^4$, although they can be detected by GW detectors, the viable parameter space is excluded by WD and NS constraints. The NS constraints are more stringent due to the higher density of NSs. For $\varepsilon \gtrsim (10^{5}~{\mathrm{GeV}})^4$, GW detectors become a valuable tool complementary to astrophysical probes. Thus, GW detectors offer a promising avenue to detect high-density DM candidates, including DQNs from dark QCD phase transitions and Fermi-balls from strong first-order phase transitions above $10^{5}$ GeV.

We also perform Fisher matrix analysis to forecast the precision of model parameters. The full analysis can be seen in the Supplementary Material. The marginalized 1-$\sigma$ uncertainty on the $i$-th parameter is estimated as the square root of the diagonal element of the covariance matrix, which is the inverse of the Fisher matrix. For example, our results indicate that for $M_{\mathrm{DM}}=10^{8}~\mathrm{g}$, $\tilde{\alpha} =10^{9.3}$ and $D = 10~\mathrm{km}$, the uncertainties are $\log_{10}(M_{\mathrm{DM}}/\mathrm{g}) = 8.0 \pm 0.19$ and $\log_{10}\tilde{\alpha}= -13.5 \pm 0.23$ at TianQin, showing the superiority of GW detectors for DM detections.

{\bf Conclusion}---
We have established a unified framework to constrain composite DM, by bridging the gap between astrophysical observations and precision interferometry. By adopting more accurate analysis of WD envelopes and incorporating expanded data, we derived a conservative bound of $\sigma_{\mathrm{DM}-A}/M_{\mathrm{DM}} \gtrsim 7.3\times 10^{-11}~\mathrm{cm}^2/\mathrm{g}$, updating previous estimates by orders of magnitude.

Complementing astrophysical bounds, we demonstrate that space-borne GW observatories, such as LISA, TianQin, and Taiji, can serve as sensitive detector by monitoring the kinematic response of their test masses to passing DM. We found that they can identify DM with masses spanning $1 \text{--} 10^{13}~\mathrm{g}$ and $\tilde{\alpha} \gtrsim 10^{5}$, given that DM interior density is high enough. Furthermore, our Fisher analysis reveals that these experiments can achieve percent-level precision in parameter estimation, highlighting their ability to provide robust measurements.

Collectively, these results place composite DM ``under siege'' from two directions. The combination of updated WD/NS constraints and the novel use of GW interferometers as particle detectors offers a powerful, multi-messenger strategy to either discover or rule out a wide class of composite candidates in the coming decade.


\let\oldaddcontentsline\addcontentsline
\renewcommand{\addcontentsline}[3]{}

{\bf Acknowledgments}---
We would like to thank Yi-Ming Hu, Daneng Yang and Sida Lu for helpful discussions. We thank Quan Chen for carefully reading our manuscript.
This work is supported by the National Natural Science Foundation of China (NNSFC) under Grant No.12475111, No.12205387, 
and the Fundamental Research Funds for the Central Universities, Sun Yat-sen University. The code that supports the findings of this article is openly available at~\cite{github_clip}.

\bibliography{reference}

\let\addcontentsline\oldaddcontentsline

\clearpage

\onecolumngrid
\begin{center}
  \textbf{\large Supplementary Material
  }\\[.2cm]
  \vspace{0.05in}
  {Siyu Jiang, Aidi Yang, Fa Peng Huang}
\end{center}

\twocolumngrid

\setcounter{equation}{0}
\setcounter{figure}{0}
\setcounter{table}{0}
\setcounter{section}{0}
\setcounter{page}{1}
\makeatletter

\setcounter{secnumdepth}{2}
\renewcommand{\thesection}{\Roman{section}}
\renewcommand{\thesubsection}{\thesection.\alph{subsection}}

\onecolumngrid

\startcontents[sections]
\tableofcontents

\section{Some details about the astrophysical constraints}
\subsection{The heat capacity for different particle species}
The heat capacity for ions, electrons, and photons is given by~\cite{Fedderke:2019jur},
\begin{align}
c_p^{\text {ion }}(T, \rho) & \equiv  \frac{5}{2 \mu_a} \sum_i \frac{X_i}{A_i}\,\,, \\
c_p^{e}(T, \rho) & \equiv \frac{\pi^2}{\mu_a \mu_e} \frac{T}{E_F}\left[1-\left(\frac{m_e}{E_F}\right)^2\right]^{-1}\,\,, \\
c_p^{\gamma}(T, \rho) & \equiv \frac{4 \pi^4}{5 \mu_a \mu_e}\left(\frac{T}{E_F}\right)^3\left[1-\left(\frac{m_e}{E_F}\right)^2\right]^{-3 / 2}\,\,,
\end{align}
where $\mu_a$ is the atomic mass unit. $\mu_e \equiv \left(\sum_i X_i Z_i / A_i\right)^{-1}$ is the mean molecular mass per electron with $X_i, Z_i$, and $A_i$ being the mass fraction, charge, and atomic mass number of ion species $i$, respectively. For WDs we assume $X_C=X_O=1/2$. $E_F=[m_e^2+ \left.\left(3 \pi^2 n_e\right)^{2 / 3}\right]^{1 / 2}$ is the electron Fermi energy and $n_e= \rho /\left(\mu_a \mu_e\right)$ is the electron number density. 

\subsection{Maximum impact factor}\label{app:impact}
Throughout the DM trajectory, it satisfies the constraint $p_{\mu}p^{\mu} = M_{\mathrm{DM}}^2$. For a DM with energy $E = \gamma M_{\mathrm{DM}} = M_{\mathrm{DM}}/\sqrt{1-v_{\mathrm{DM}}^2}$, velocity $v_{\mathrm{DM}}$, conserved angular momentum $L = \gamma M_{\mathrm{DM}} b v_{\mathrm{DM}}$ with $b$ being the impact factor, we get the following relation~\cite{Gresham:2022biw},
\begin{equation}
\left(\frac{d r}{d \tau}\right)^2=\left(\frac{E-V_{\mathrm{eff},+}}{M_{\mathrm{DM}}}\right)\left(\frac{E-V_{\mathrm{eff},-}}{M_{\mathrm{DM}}}\right)
\end{equation}
with
\begin{align}
\frac{V_{\mathrm{eff}, \pm}}{M_{\mathrm{DM}}}=&-\frac{G M_\mathrm{CO} \tilde{\alpha} e^{-m_\phi r }}{r} \pm \sqrt{\left(1-\frac{2 G M_\mathrm{CO}}{r}\right)\left(\frac{(L / M_{\mathrm{DM}})^2}{r^2}+1\right)} \,\,,
\end{align}
where $\tilde{\alpha} = \delta_{\mathrm{DM}} \delta_{\mathrm{SM}}$.
The maximum impact factor can be found by using the following conditions~\cite{Gresham:2022biw}
\begin{equation}
\left.V_{\mathrm{eff},+}^{\prime}\left(r\right)\right|_{L=\gamma M_{\mathrm{DM}} b_{\max} v_{\mathrm{DM}}}=0\,\,,\\
\left. V_{\mathrm{eff},+}\left(r\right)\right|_{L=\gamma M_{\mathrm{DM}} b_{\max} v_{\mathrm{DM}}}=\gamma M_{\mathrm{DM}}\,\,.
\end{equation}
Using $v_{\mathrm{DM}}^2 \ll \frac{G M_{\mathrm{CO}}}{R_{\mathrm{CO}}} \ll 1$ and $\lambda \gg R_{\mathrm{CO}}$, the analytic approximation for $b_{\text {max }}$ is given by \begin{equation}
b_{\max }=\min \left(b_{\max , \text { in }}, b_{\max , \text { out }}\right)\,\,,
\end{equation}
with
\begin{equation}
    b_{\max, \text { in}} \approx  \frac{R_\mathrm{CO}}{v_\mathrm{DM}} \sqrt{\frac{\frac{2 G M_\mathrm{CO}}{R_\mathrm{CO}}\left(1+\tilde{\alpha} \right)+\left(\frac{G M_\mathrm{CO}}{R_\mathrm{CO}} \tilde{\alpha} \right)^2}{\left(1-\frac{2 G M_\mathrm{CO}}{R_\mathrm{CO}}\right)}}\,\,,
\end{equation}
and
\begin{equation}
b_{\max , \text { out }} \approx \sqrt{\lambda x\left(\frac{2 G M_\mathrm{CO}}{v_\mathrm{DM}^2}+\lambda x\right)}, \quad x \approx \log \left(\frac{\tilde{\alpha}}{\frac{\lambda v_\mathrm{DM}^2}{G M_\mathrm{CO}}+\frac{1}{\log \tilde{\alpha}}}\right)\,\,.
\end{equation}
When $\tilde{\alpha} \rightarrow 0$, the maximum impact factor is
\begin{equation}
    b_{\mathrm{max}} = \frac{R_{\mathrm{CO}}}{v_{\mathrm{DM}}} \sqrt{\frac{v_{\mathrm{esc}}^2}{1-v_{\mathrm{esc}}^2}}\,\,,
\end{equation}
which reproduces the formula without the Yukawa interaction.

\subsection{The width of a non-degenerate WD envelope}
The density $\rho_*$ and the temperature $T_*$  at the transition point can be derived
by equating the electron pressure of the degenerate layer
to that of the non-degenerate layer~\cite{Shapiro:1983du},
\begin{eqnarray}
\frac{\rho_* k T_*}{\mu_e \mu_a}=1.0 \times 10^{13}\left(\frac{\rho_*}{\mu_e}\right)^{5 / 3}\,\,,
\end{eqnarray}
from which one gets,
\begin{eqnarray}\label{rhoTs}
\rho_*=\left(2.4 \times 10^{-8} \mathrm{~g}\cdot\mathrm{~cm}^{-3}\right) \mu_e T_*^{3 / 2}\,\,.
\end{eqnarray}
The photon diffusion equation is $L_{\mathrm{WD}}=-4 \pi r^2 \frac{c}{3 \kappa \rho} \frac{d}{d r}\left(a T^4\right)$ with $a$ being the radiation constant and $L_{\mathrm{WD}}$ the luminosity. $\kappa$ is the 
opacity of the stellar material and can be approximated by Kramer's opacity $\kappa=\kappa_0 \rho T^{-3.5}$ with 
\begin{eqnarray}
\kappa_0=4.34 \times 10^{24} Z(1+X)~\mathrm{cm}^2 \cdot\mathrm{~g}^{-1} \,\,.
\end{eqnarray}
where $X$ is the mass fraction of hydrogen and $Z$ is the mass fraction of heavy elements other than hydrogen and helium. The photon diffusion equation gives,
\begin{eqnarray}\label{dTdr}
\frac{d T}{d r}=-\frac{3}{4 a c} \frac{\kappa \rho}{T^3} \frac{L_{\mathrm{WD}}}{4 \pi r^2} .
\end{eqnarray}

By combining the photon diffusion equation and the equation of hydrostatic equilibrium $\frac{d P}{d r}=-\frac{G M_{\mathrm{WD}} }{r^2}\rho$, the pressure $P$ follows the following equation~\cite{Shapiro:1983du},
\begin{eqnarray}
P d P=\frac{4 a c}{3} \frac{4 \pi G M_{\mathrm{WD}}}{\kappa_0 L_{\mathrm{WD}}} \frac{k}{\bar{\mu} \mu_a} T^{7.5} d T\,\,.
\end{eqnarray}
After integration and replacing $P$ by $\rho$ from the equation of state, 
\begin{eqnarray}\label{rhoTs2}
\rho=\left(\frac{8 a c}{25.5} \frac{4 \pi G M_{\mathrm{WD}}}{\kappa_0 L_{\mathrm{WD}}} \frac{\bar{\mu} \mu_a}{k}\right)^{1 / 2} T^{3.25}\,\,.
\end{eqnarray}
Combining Eqs.~\eqref{rhoTs} and \eqref{rhoTs2}, the luminosity $L_{\mathrm{WD}}$ can be expressed as, 
\begin{equation}\label{luminosity}
    L_{\mathrm{WD}}=(5.7\times10^5~\mathrm{erg}\cdot\mathrm{s}^{-1})\times \frac{\bar{\mu}}{\mu_e^2}\frac{1}{Z(1+X)}\frac{M_{\mathrm{WD}}}{M_\odot}\left(\frac{T_*}{1~\mathrm{K}}\right)^{3.5}\,\,,
\end{equation}
where $\bar{\mu}\equiv \left(\sum_i X_i (1+Z_i) / A_i\right)^{-1}$ is the mean molecular weight. For WDs, $L_{\mathrm{WD}}$ can be derived from the cooling rate of the total thermal energy~\cite{Shapiro:1983du},
\begin{align}\label{luminosity2}
    L_{\mathrm{WD}}  
    \approx (3.14\times 10^{24}~\mathrm{erg}\cdot\mathrm{s}^{-1}) \times 
    \left(\frac{1}{\bar{\mu}}-\frac{1}{\mu_e}\right)\left(\frac{M_{\mathrm{WD}}}{M_\odot}\right)\left(\frac{1~\mathrm{Gyr}}{\tau_{\mathrm{WD}}}\right)\left(\frac{T_*}{1~\mathrm{K}}\right)\,\,.
\end{align}
Combining Eqs.~\eqref{luminosity} and ~\eqref{luminosity2} one can get
\begin{align}
    \left(\frac{T_*}{1~\mathrm{K}}\right)^{2.5} \simeq 5.5\times 10^{18}~Z(1+X)\frac{\mu_e}{\bar{\mu}}\left(\frac{\mu_e}{\bar{\mu}}-1\right)\left(\frac{1~\mathrm{Gyr}}{\tau_{\mathrm{WD}}}\right)\,\,.
\end{align}
After obtaining $T_*$ one can eliminate $\rho$ in Eq.~\eqref{dTdr} by using Eqs.~\eqref{rhoTs2} and $\kappa=\kappa_0 \rho T^{-3.5}$, and integrate to get 
\begin{equation}
    T_*=\frac{1}{4.25}\frac{\bar{\mu} \mu_a}{k}\frac{GM_{\mathrm{WD}}}{R_{\mathrm{WD}}}\left(\frac{R_{\mathrm{WD}}}{R_*}-1\right)\,\,,
\end{equation}
then the width of the non-degenerate WD envelope is obtained by using $R_{\mathrm{env}} = R_{\mathrm{WD}} - R_{*}$.

\section{More details on the signals induced by DM in GW detectors}
\subsection{Signal and noise PSD for $X$, $\alpha$ and $\zeta$ combinations}\label{app:fullsignal}
\begin{table}[h]
    \centering
\begin{tabular}{|c|c|c|c|}
\hline
Parameter & LISA & TianQin & Taiji  \\
\hline
Arm length $L$ ($10^9$~m) & 2.5 & 0.17 & 3  \\
$P_{\mathrm{oms}}$ ($10^{-12}$~m) & 15 & 1 & 8  \\
$P_{\mathrm{acc}}$ ($10^{-15}$~m $\cdot$ s$^{-2}$) & 3 & 1 & 3  \\
$T_{\mathrm{obs}}$ (yr) & 4.5 & 2.5 & 5  \\
Frequency range (Hz) & [$10^{-4}$, 1] & [$10^{-4}$, 1] & [$10^{-4}$, 1]  \\
\hline
\end{tabular}
\caption{The parameters of different GW detectors.}
\label{tab:detector}
    
\end{table}

The Fourier transformation of the velocity of the satellite \eqref{vt} in the main text is,
\begin{equation}
\frac{\tilde{\vec{v}}}{c}(\omega) = \frac{2\tilde{G}M_{\mathrm{DM}}}{cv_{\mathrm{DM}}^2} \begin{bmatrix}
\mathrm{i}K_1(\omega D/v_{\mathrm{DM}}) \\
K_0(\omega D/v_{\mathrm{DM}}) \\
0
\end{bmatrix}\,\,,
\end{equation}
with $\omega = 2\pi f$. Then the Fourier transformation of Eq.~\eqref{Xt} in the main text can be obtained,
\begin{align}
\tilde{X}(\omega) = &\sqrt{\frac{2}{\pi}} \left(1 - \mathrm{e}^{4i\omega L/c}\right) \frac{\tilde{G}M_{\mathrm{DM}}}{cv_{\mathrm{DM}}^2} \sin\vartheta \times \left[K_0\left(\frac{D\omega}{v_{\mathrm{DM}}}\right) \sin\varphi - iK_1\left(\frac{D\omega}{v_{\mathrm{DM}}}\right) \cos\varphi\right] + \nonumber\\
&\sqrt{\frac{8}{\pi}} \left(\mathrm{e}^{3i\omega L/c} - \mathrm{e}^{i\omega L/c}\right) \frac{\tilde{G}M_{\mathrm{DM}}}{cv_{\mathrm{DM}}^2} \sin\vartheta \times \left[K_0\left(\frac{D'\omega}{v_{\mathrm{DM}}}\right) \sin\varphi - iK_1\left(\frac{D'\omega}{v_{\mathrm{DM}}}\right) \cos\varphi\right]\,\,.
\end{align}
Then the signal PSD reads,
\begin{align}\label{PX}
P(\omega)  =&\left\langle \left| \tilde{X}(\omega)\right|^2\right\rangle 
=\frac{1}{4 \pi} \int \mathrm{d} \vartheta \sin \vartheta \int \mathrm{~d} \varphi|\tilde{X}(\omega)|^2 \\
= &\frac{32}{3\pi} \left(\frac{\tilde{G}M_{\mathrm{DM}}}{cv_{\mathrm{DM}}^{2}}\right)^{2}\sin^{2}\left(\omega L/c\right) \times \left\{ \left[K_{0}\left(\frac{D\omega}{v_{\mathrm{DM}}}\right)\cos(\omega L/c) - K_{0}\left(\frac{D'\omega}{v_{\mathrm{DM}}}\right)\right]^2+ \right. \nonumber\\
&\left. \left[K_{1}\left(\frac{D\omega}{v_{\mathrm{DM}}}\right)\cos(\omega L/c) - K_{1}\left(\frac{D'\omega}{v_{\mathrm{DM}}}\right)\right]^2\right\} \,\,.
\end{align}
It can be seen that in the limit $D \ll L$, $|\vec{v}_2|, |\vec{v}_3| \ll |\vec{v}_1|$, and we can neglect the terms including $D'$ in the bracket. The equation then reduces to the results given by \cite{Jaeckel:2020mqa}.

The noise PSD for the X channel is given by expressions of the form~\cite{Hartwig:2023pft}
\begin{align}
S_{X}(f) = &16 \sin^{2}\left(\frac{2 \pi f L}{c}\right) \biggl\{\left[3 + \cos\left(\frac{4 \pi f L}{c}\right)\right] S_{\mathrm{acc}}(f) +  S_{\mathrm{oms}}(f) \biggl\} \,\,,
\end{align}
where $S_{\mathrm{oms}}(f)$ and $S_{\mathrm{acc}}(f)$ denote the PSDs of the optical measurement system (OMS) and acceleration noises, respectively. The noise PSDs for LISA, TianQin, and Taiji are given by~\cite{babak2021lisasensitivitysnrcalculations,Flauger:2020qyi},
\begin{align}
S_{\mathrm{oms}}(f)=&\left(\frac{2 \pi f P_{\mathrm{oms}} }{c} \right)^{2}\left[1+\left(\frac{2 \times 10^{-3} \mathrm{~Hz}}{f}\right)^{4}\right] \mathrm{Hz}^{-1}\\
S_{\mathrm{acc}}(f)=&\left(\frac{P_{\mathrm{acc}}}{2 \pi f c}\right)^{2}\left[1+\left(\frac{0.4 \times 10^{-3} \mathrm{~Hz}}{f}\right)^{2}\right]\times \left[1+\left(\frac{f}{8 \times 10^{-3} \mathrm{~Hz}}\right)^{4}\right] \mathrm{Hz}^{-1}\,\,,
\end{align}
where the values of $P_{\mathrm{oms}}$ and $P_{\mathrm{acc}}$ are given in Tab.~\ref{tab:detector}.

For BBO, the noise PSDs are given by~\cite{Corbin:2005ny}
\begin{equation}
S_{\mathrm{oms}}^{\mathrm{BBO}}(f)=\frac{2.0 \times 10^{-34}\mathrm{~m}^2}{3 L^2}  \mathrm{~Hz}^{-1}\,\,,
\end{equation}
and
\begin{equation}
S_{\mathrm{acc}}^{\mathrm{BBO}}(f)=\frac{4.5 \times 10^{-34} \mathrm{~m}^2 \cdot \mathrm{~Hz}^4}{(2 \pi f)^4(3 L)^2} \mathrm{~Hz}^{-1}\,\,,
\end{equation}
with $L = 5\times 10^4~\mathrm{km}$ and $T_{\mathrm{obs}} = 4~\mathrm{yr}$ for BBO.

\begin{figure}[!ht]
    \centering
    \includegraphics[width=0.7\linewidth]{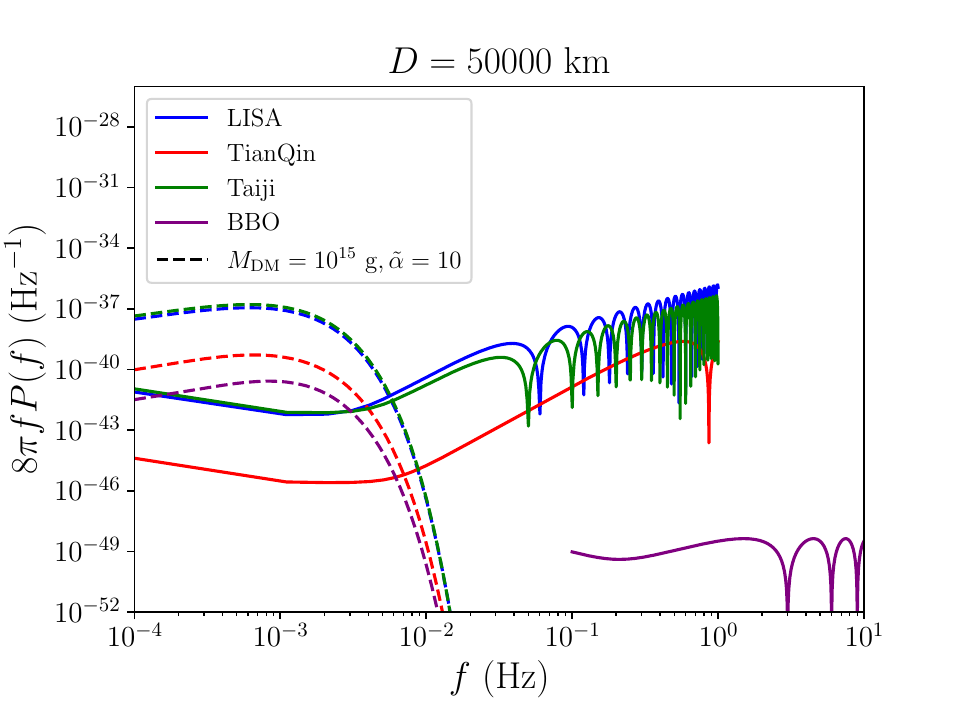}
    \caption{The signal PSD (dashed lines) of composite DM  on the GW detectors in the X channel and the corresponding noise PSD (solid lines).}
    \label{DM_PSD}
\end{figure}
In Fig.~\ref{DM_PSD} we show the DM signals and the corresponding noise PSD for LISA~\cite{amaroseoane2017laserinterferometerspaceantenna}, TianQin~\cite{TianQin:2015yph}, Taiji~\cite{Hu:2017mde}, and BBO~\cite{Corbin:2005ny}. We fix impact factor $D = 50000~\mathrm{km}$.

Similar to the Michelson-like combination X, the Sagnac combination $\alpha$ is given by
\begin{align}
    &\alpha(t) = U_1(t) + V_1(t) + U_3(t-\frac{L}{c}) +V_2(t-\frac{L}{c}) + U_2(t-\frac{2L}{c}) + V_3(t-\frac{L}{c}) \nonumber \\
    &= -\vec{n}_1 \cdot \frac{\vec{v}(t,D) - \vec{v}(t-3L/c,D)}{c}  -3\vec{n}_1 \cdot \frac{\vec{v}(t-2L/c,D') - \vec{v}(t-L/c,D')}{c}\,\,,
\end{align}
and the fully symmetric Sagnac combination $\zeta$ is
\begin{align}
    \zeta(t) &= U_1(t-\frac{L}{c}) + V_1(t-\frac{L}{c}) + U_2(t-\frac{L}{c}) + V_2(t-\frac{L}{c}) + U_3(t-\frac{L}{c}) + V_3(t-\frac{L}{c}) \nonumber \\
    & = -\vec{n}_1 \cdot \left[ \frac{\vec{v}(t-L/c,D)-\vec{v}(t-2L/c,D)}{c} +  \frac{\vec{v}(t-2L/c,D')-\vec{v}(t-L/c,D')}{c}\right]\,\,.
\end{align}
After Fourier transformation, we have
\begin{align}
\tilde{\alpha}(\omega) &= \sqrt{\frac{2}{\pi}} \left(1 - \mathrm{e}^{3i\omega L/c}\right) \frac{\tilde{G}M_{\mathrm{DM}}}{cv_{\mathrm{DM}}^2} \sin\vartheta \times \left[K_0\left(\frac{D\omega}{v_{\mathrm{DM}}}\right) \sin\varphi - iK_1\left(\frac{D\omega}{v_{\mathrm{DM}}}\right) \cos\varphi\right] + \nonumber\\
&3\sqrt{\frac{2}{\pi}} \left(\mathrm{e}^{2i\omega L/c} - \mathrm{e}^{i\omega L/c}\right) \frac{\tilde{G}M_{\mathrm{DM}}}{cv_{\mathrm{DM}}^2} \sin\vartheta \times \left[K_0\left(\frac{D'\omega}{v_{\mathrm{DM}}}\right) \sin\varphi - iK_1\left(\frac{D'\omega}{v_{\mathrm{DM}}}\right) \cos\varphi\right]\,\,,
\end{align}
which leads to the PSD in the $\alpha$ channel,
\begin{align}
  P_\alpha (\omega) = \left\langle \left| \tilde{\alpha}(\omega)\right|^2\right\rangle  = &\frac{8}{3\pi} \left(\frac{\tilde{G}M_{\mathrm{DM}}}{cv_{\mathrm{DM}}^{2}}\right)^{2}\sin^{2}\left(\frac{\omega L}{2c}\right) \times \left\{ \left[K_{0}\left(\frac{D\omega}{v_{\mathrm{DM}}}\right)\left(1+2\cos\left(\frac{\omega L}{c}\right)\right) - 3K_{0}\left(\frac{D'\omega}{v_{\mathrm{DM}}}\right)\right]^2+ \right. \nonumber\\
&\left. \left[K_{1}\left(\frac{D\omega}{v_{\mathrm{DM}}}\right)\left(1+2\cos\left(\frac{\omega L}{c}\right)\right) - 3K_{1}\left(\frac{D'\omega}{v_{\mathrm{DM}}}\right)\right]^2\right\} \,\,,
\end{align}
and
\begin{align}
\tilde{\zeta}(\omega) &= \sqrt{\frac{2}{\pi}} \left(\mathrm{e}^{i\omega L/c}- \mathrm{e}^{2i\omega L/c}\right) \frac{\tilde{G}M_{\mathrm{DM}}}{cv_{\mathrm{DM}}^2} \sin\vartheta \times \left[K_0\left(\frac{D\omega}{v_{\mathrm{DM}}}\right) \sin\varphi - iK_1\left(\frac{D\omega}{v_{\mathrm{DM}}}\right) \cos\varphi\right] + \nonumber\\
&\sqrt{\frac{2}{\pi}} \left(\mathrm{e}^{2i\omega L/c} - \mathrm{e}^{i\omega L/c}\right) \frac{\tilde{G}M_{\mathrm{DM}}}{cv_{\mathrm{DM}}^2} \sin\vartheta \times \left[K_0\left(\frac{D'\omega}{v_{\mathrm{DM}}}\right) \sin\varphi - iK_1\left(\frac{D'\omega}{v_{\mathrm{DM}}}\right) \cos\varphi\right]\,\,,
\end{align}
which leads to the PSD in the $\zeta$ channel,
\begin{align}
  P_\zeta (\omega) = \left\langle \left| \tilde{\zeta}(\omega)\right|^2\right\rangle  = &\frac{8}{3\pi} \left(\frac{\tilde{G}M_{\mathrm{DM}}}{cv_{\mathrm{DM}}^{2}}\right)^{2}\sin^{2}\left(\frac{\omega L}{2c}\right) \times \left\{ \left[K_{0}\left(\frac{D\omega}{v_{\mathrm{DM}}}\right) - K_{0}\left(\frac{D'\omega}{v_{\mathrm{DM}}}\right)\right]^2+ \right. \nonumber\\
&\left. \left[K_{1}\left(\frac{D\omega}{v_{\mathrm{DM}}}\right) - K_{1}\left(\frac{D'\omega}{v_{\mathrm{DM}}}\right)\right]^2\right\} \,\,.
\end{align}

The corresponding noise PSD for $\alpha$ and $\zeta$ channels reads,
\begin{align}
S_\alpha(f) & =8\left[2 \sin ^2(\pi f L)+\sin ^2(3 \pi f L)\right] S_{\mathrm{acc}}(f)+6 S_{\mathrm{oms}}(f)\,\,, \\
S_\zeta(f) & =6\left[4 \sin ^2(\pi f L) S_{\mathrm{acc}}(f)+S_{\mathrm{oms}}(f)\right]\,\,.
\end{align}

\subsection{Expected number of DM passes through the detectors for fixed SNR}\label{app:eventrate}

\begin{figure}[!ht]
    \centering
    \includegraphics[width=0.7\linewidth]{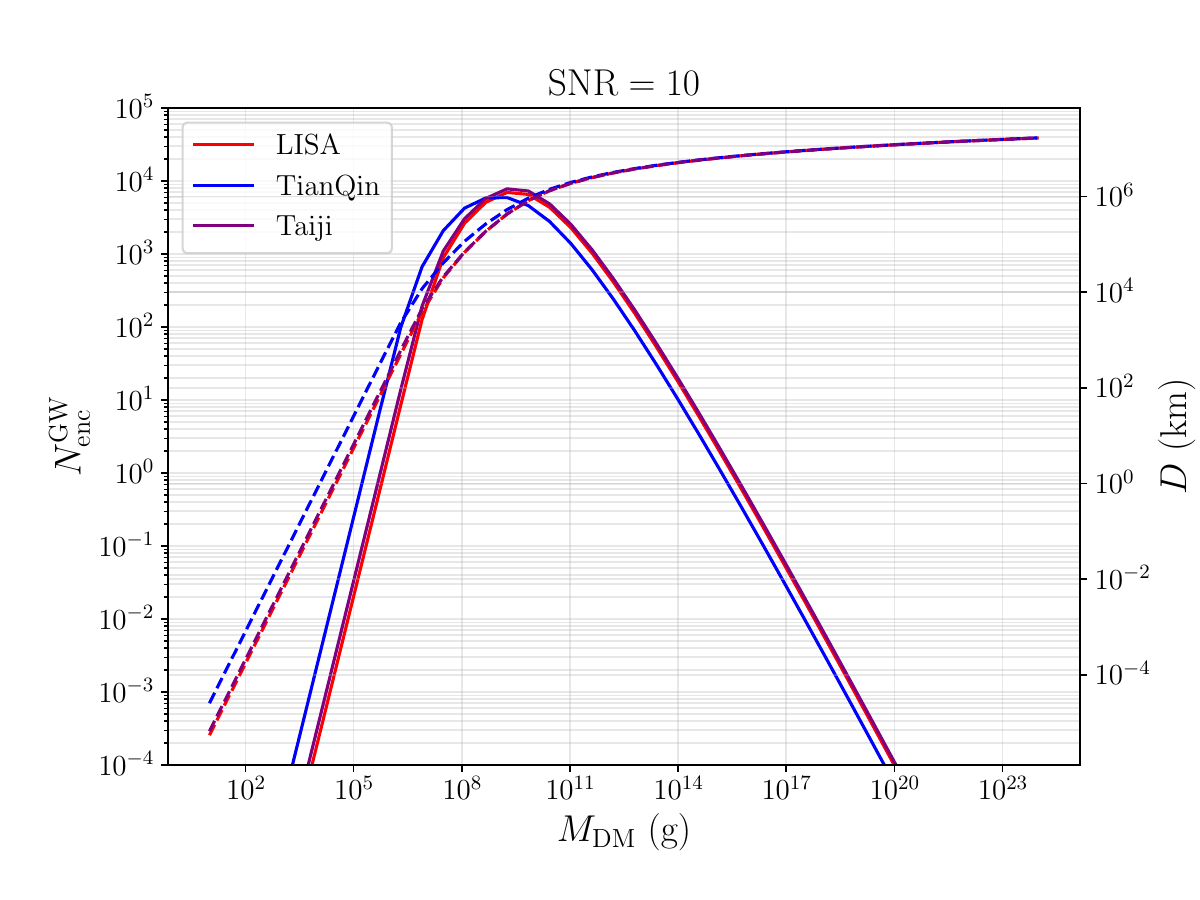}
    \caption{Expected number of DM passes through the detectors (solid lines) and impact factor (dashed lines) of composite DM as functions of DM mass in the X channel.}
    \label{DM_SNR10}
\end{figure}

In Fig.~\ref{DM_SNR10}, we show the expected number of DM passes through the detectors $N_{\mathrm{enc}}^{\mathrm{GW}}=\left(f_{\mathrm{DM}} \pi D^2 \rho_{\mathrm{DM}} v_{\mathrm{DM}}/M_{\mathrm{DM}}\right) T_{\mathrm{obs}}^i$ for fixed SNR=10 for LISA, Taiji, and TianQin in the X channel, with $\tilde{\alpha} = 10$. It can be seen that the expected event number is lower when the DM mass is smaller or larger. This can be understood as follows: when DM mass is smaller, the DM number density is large, and the averaged distance between DM and GW detector is small, $D \ll L$. In this case, from the Taylor expansion of Eq.~\eqref{PX} we get $ P(\omega) \propto M_{\mathrm{DM}}^2 / D^2 \simeq \text { constant }$, then $ N_{\mathrm{enc}}^{\mathrm{GW}} \propto M_{\mathrm{DM}}$. If the DM mass is large enough, the averaged distance satisfies $D \gg L$. After the Taylor expansion of Eq.~\eqref{PX} at large $D \omega/ v_{\mathrm{DM}}$, one can obtain $P(w) \propto \frac{M_{\mathrm{DM}}^2 \mathrm{e}^{-2 D w / v_{\mathrm{DM}}}}{D}$ and $N_{\mathrm{enc}}^{\mathrm{GW}} \propto \frac{\left(\log M_{\mathrm{DM}}\right)^2}{M_{\mathrm{DM}}}$. These evaluations are both consistent with the behavior of $N_{\mathrm{enc}}^{\mathrm{GW}}$ in Fig.~\ref{DM_SNR10}.

\subsection{Fisher matrix analysis and model parameter reconstruction}\label{app:fullfisher}

The Fisher matrix approach provides a method for estimating parameter uncertainties by approximating the likelihood function as a Gaussian distribution:
\begin{equation}
\mathcal{L}(\vec{\theta})\propto\exp(-\vec{\theta}^T\cdot\mathcal{C}^{-1}\cdot\vec{\theta}/2)\,\,,
\end{equation}
where $\vec{\theta}$ is a series of model parameters.

The covariance matrix $\mathcal{C}$ is obtained by inverting the Fisher matrix $\mathcal{F}$. The elements of $\mathcal{F}$ are derived from the model of the signal $h(\vec{\theta})$, and are expressed as:
\begin{equation}
\mathcal{F}_{ij}=\sum_{k=X, \alpha, \zeta}\langle\partial_{\theta_i}h^k|\partial_{\theta_j}h^k\rangle\,\,.
\end{equation}
In this work, we simply set the signal $h$ to be $\sqrt{P(f)}$. The inner product is defined as follows:
\begin{equation}
\langle a|b\rangle=4\int\limits_0^\infty\mathrm{d}f\frac{\Re(a(f)b^*(f))}{S_n(f)}
\end{equation}
with $S_{n}(f)$
being the instrument noise PSD. The numerical challenge is connected to the inversion of the Fisher matrix. This issue comes from the large range of numerical values, e.g., large values of DM mass and small values of $\alpha_X$. So we choose the parameter set to be $\left\{\log_{10}(M_\mathrm{DM}/\mathrm{g}), \log_{10}\tilde{\alpha}, \log_{10}(D/\mathrm{km}), v_{\mathrm{DM}}(\mathrm{km/s}) \right\}$. 

\begin{figure}[!ht]
    \centering
    \includegraphics[width=0.8\linewidth]{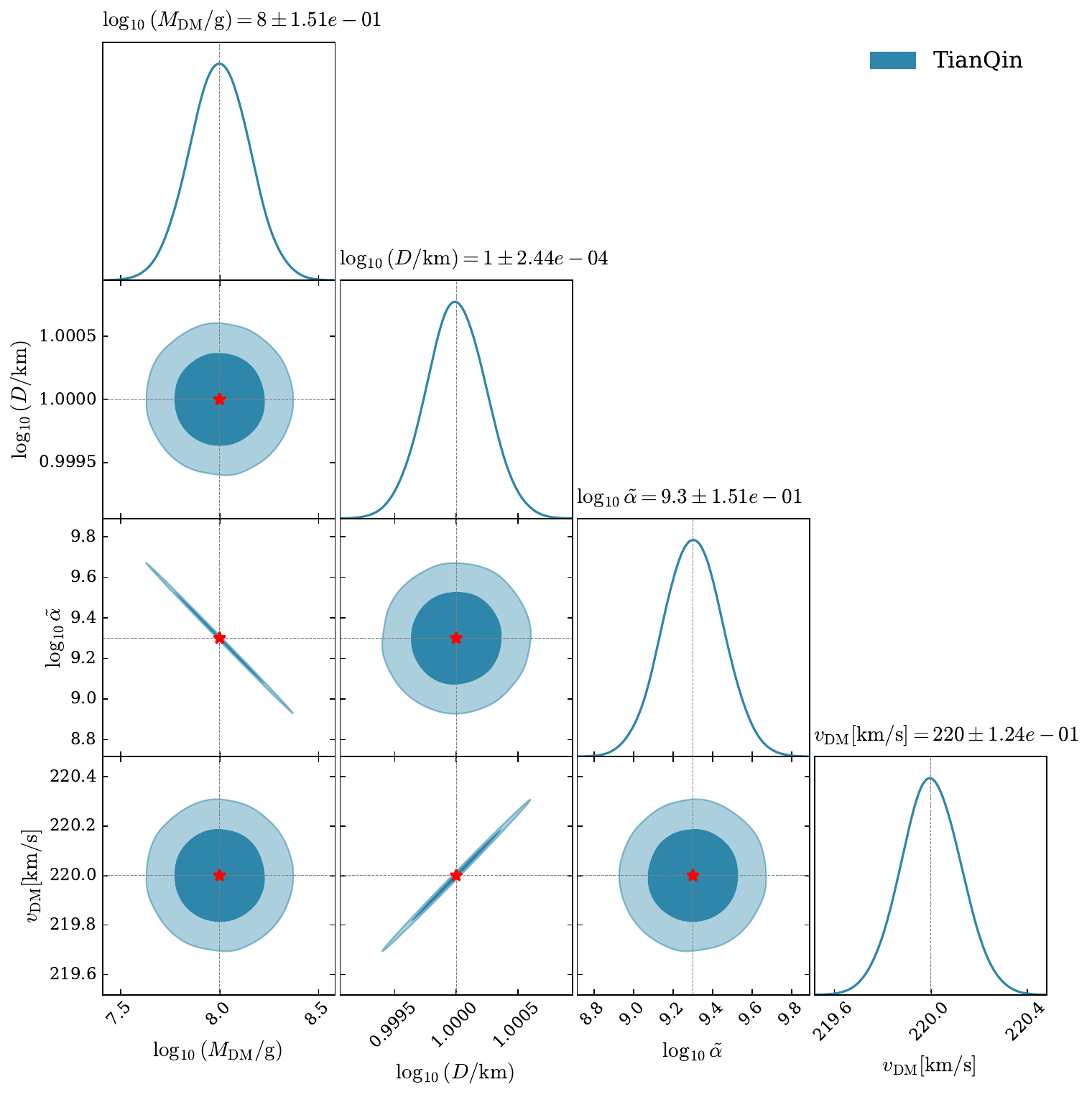}
    \caption{Triangle plot  of the Fisher analysis  for a signal at TianQin induced by composite DM.}
    \label{fig:tianqin}
\end{figure}

\begin{figure}[!ht]
    \centering
    \includegraphics[width=0.8\linewidth]{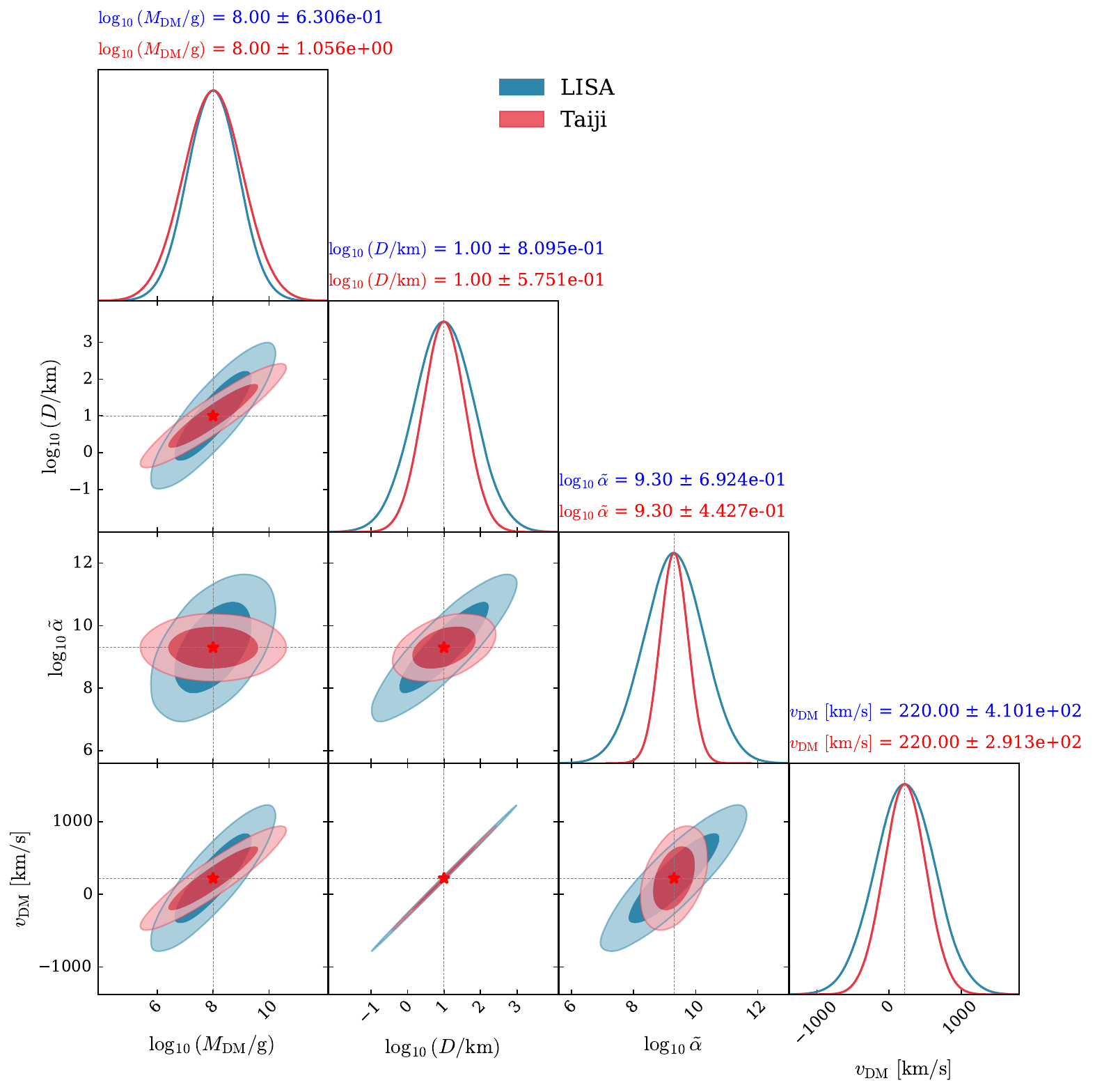}
    \caption{Triangle plot  of the Fisher analysis  for a signal at LISA and Taiji induced by composite DM.}
    \label{LISATaiji}
\end{figure}
Fig.~\ref{fig:tianqin} and Fig.~\ref{LISATaiji} present the projected $68\%$ and $95\%$ confidence level (CL) contours for a DM signal for TianQin and LISA+Taiji, respectively. The analysis assumes a fiducial parameter set of $\left\{\log_{10}(M_\mathrm{DM}/\mathrm{g})=8, \log_{10}\tilde{\alpha}=9.3,   \log_{10}(D/\mathrm{km})=1,  v_{\mathrm{DM}}(\mathrm{km/s})=220 \right\}$. In this visualization, the diagonal panels display the one-dimensional marginalized posterior distributions, while the off-diagonal panels reveal the two-dimensional covariance ellipses. The numerical values above each column indicate the inferred means and $1\sigma$ uncertainties derived from the Fisher matrix analysis.

The results demonstrate that TianQin can constrain the model parameters with remarkable precision. Specifically, the DM mass $M_\mathrm{DM}$ is determined with a relative error of approximately $4\%$ (at $68\%$ CL). This tight constraint is anticipated, given the high sensitivity of the GW signal to the underlying DM mass.  The distance parameter $D/\mathrm{km}$ is constrained with a relative precision of $\sim 0.025\%$. Similarly, the DM velocity $v_{\mathrm{DM}}$ is determined with a relative error of $\sim 0.06\%$.  This sub-percent precision indicates that TianQin is highly sensitive to the spectral features governed by the velocity and the distance. In contrast, $\tilde{\alpha}$ exhibits a larger uncertainty. The comparatively looser constraint on $\log_{10}\tilde{\alpha}$—despite the high precision on $\log_{10}M_\mathrm{DM}$—suggests that the intrinsic coupling strength suffers from partial degeneracies with the DM mass. A notable correlation is observed between $\log_{10}\tilde{\alpha}$ and  $\log_{10}M_\mathrm{DM}$, as evidenced by the upward-left tilt (negative correlation) of the corresponding contour.  Consequently, a source placed at a larger  $\log_{10}M_\mathrm{DM}$ can produce the same observed signal strength if it possesses a lower $\log_{10}\tilde{\alpha}$. We also observe the degeneracy between the velocity $v_{\mathrm{DM}}$ and the distance $\log_{10}(D/\mathrm{km})$, as evidenced by the elongated contours in the corresponding subplot (upward-right). 

We further explore the parameter reconstruction capabilities of LISA (steel blue contours) and Taiji (red contours) in Fig.~\ref{LISATaiji}. We have separated these results from TianQin due to the distinct PSDs. Overall, the comparative results for both detectors exhibit similar qualitative behaviors, although the resulting uncertainties are generally larger than those obtained for TianQin. LISA  exhibits visibly tighter confidence regions across the entire parameter space relative to Taiji. This difference is likely due to the specific signal falling closer to the optimal sensitivity band of LISA.  The strong degeneracy between $\log_{10}\tilde{\alpha}$ and $\log_{10}(D/\mathrm{km})$ persists in both cases. Other parameter pairs also display distinct elliptical shapes, indicating their correlations.
\end{document}